\documentstyle[preprint,aps]{revtex}
\begin{document}
\title{
Hidden Duality and associated instabilities 
of Tomonaga-Luttinger Liquid on Lattice}
\author{Z. N. C. Ha\footnote{zha@magnet.fsu.edu 
(Email); 904-644-5038 (Fax).}}
\address{
National High Magnetic Field Laboratory,
Florida State University, Tallahassee, FL 32306, U.S.A.
}
\date{July 24, 1997}
\maketitle
\begin{abstract}
Hidden duality and associated instabilities of the spinless
Luttinger liquid on lattice are reported. The local quantum
fluctuations and the long-distance chiral modes compete and as a
result produce a hierarchy of exotic charge/density instabilities.
Explicit bosonic quantum operators for the local density
fluctuations are constructed and are used to make identification of
the Luttinger liquid with the classical 2D Coulomb gas with
$\theta$-term and with the rich hidden duality.
\end{abstract}
\newpage
\tightenlines
I report novel instabilities and hidden duality structure of the
spinless Luttinger liquid on lattice with anomalous {\it local}
quantum fractional statistical fluctuations which, I argue, are
induced by the possible multi-particle umklapp and pairing
processes at rational densities. The local parameter is, in
general, different from the usual charge stiffness $K$ which
characterizes the {\it long-distance} physics.  It is shown that
the interplay between the {\it short-} and {\it long-distance}
physics gives rise to new exotic charge/density instabilities which
expose the hidden duality of the Luttinger liquid on lattice.

First, I need to consider two conjugate phase fields $\theta(x)$
and $\phi(x)$. The phase $\theta(x)$ essentially defines a field
that measures the density modulations, and it acquires phase $\pm
2\pi$ going from one to a neighboring particle. The canonically
conjugate phase field $\phi(x)$ is associated with the $U(1)$
charge degrees of freedom such that $[\phi(x),\theta(x')] = i\pi
\mbox{sgn}(x-x')$. Explicit constructions of the two operators are
straightforward via Fourier transform. Usually, the renormalization
coupling constant $e^{-2\varphi}$ ($= 2\pi K$) is introduced to
code the effects of quantum fluctuation [1].

In the Luttinger liquid universality class
it is always possible to find the
right and left eigenmodes which carry, in general, fractional statistics.
In order to show this more explicitly I use the following
right and left Mandelstam modes [2]
\begin{equation}
\Psi_R^\dagger(x) = e^{i\phi(x)} e^{i \beta\theta(x)}; \quad \quad
\Psi_L^\dagger(x) = e^{i\phi(x)} e^{-i \beta\theta(x)}.
\end{equation}
The time-dependent correlation function for large $x$ and $t$ is
given by
\begin{equation}
\langle \Psi_R^\dagger(x,t)\Psi_R(0,0) \rangle
\propto  {1\over (x-v_s t)^{2x_R} (x+v_s t)^{2x_L}},
\end{equation}
where $x_{R,L} = (2\beta \exp(\varphi) \pm \exp(-\varphi))^2/4$. If
$2\beta = \exp(-2\varphi)$ then the correlation function involves
either the right- or the left-movers only. Therefore, the
Mandelstam modes with $2\beta = \exp(-2\varphi)$ can be regarded as
the FQS-carrying {\it long-distance} chiral eigenmodes of the
Luttinger liquid.

I conjecture that there are two parameters for describing the
Luttinger liquid on lattice. One is the long-distance charge
stiffness $K (=\exp(-2\varphi)/2\pi)$ previously discussed, and the
other the local FQHE-like parameter which is presumably determined
by the filling fraction and {\it allowed} local interactions such
as the umklapp processes.

Now, consider perturbing the Luttinger liquid with the density
fluctuations containing the following
\begin{equation}
\Psi^{\dagger m}_{\lambda,n} = e^{i n \theta(x)}e^{i m\phi(x)}
e^{i m\lambda \theta(x)},
\end{equation}
where $m$ is integer content of the $U(1)$ charge and $n$ integer
index for the low-energy sectors.  The parameters $m$, $n$, and
$\lambda$ are chosen such that the overall operator be bosonic and
the resulting damping term equal to constant (allowed umklapp
condition).  The scaling dimension,
\begin{equation}
x^\lambda_{n,m} = {1\over 4\pi K} (n + \lambda m)^2
+ \pi K m^2,
\end{equation}
is invariant under the following duality $\hat{D}$ and periodicity
$\hat{T}$ transformations [3]
\begin{eqnarray}
&\hat{D}:& \eta \rightarrow  1/\eta; \quad\quad
(n,m) \rightarrow (-m,n), \label{dual} \\
&\hat{T}:& \eta \rightarrow  \eta + i; \quad\quad
(n,m) \rightarrow (n-m,m),
\end{eqnarray}
where ${\hat D}^2 = 1$ and
$\eta$ is a complex parameter defined as $2\pi K + i\lambda$.
This duality generalizes the well-known duality for $\lambda = 0$.

This work is supported by DOE grant 5024-528-23.
\vskip 0.5in
\centerline{REFERENCES}
\begin{enumerate}
\item F. D. M. Haldane, Phys. Rev. {\bf B25}, 4925 (1982).
\item S. Mandelstam, Phys. Rev. {\bf D11}, 3026 (1975).
\item J. L. Cardy, Nucl. Phys. {\bf B205},  17 (1982).
\end{enumerate}
\end{document}